\documentclass[twocolumn,twocolappendix]{aastex63}

\usepackage{amsmath}

\definecolor{dgreen}{RGB}{26,148,49}

\shorttitle{Near-IR EBL Fluctuations on Nonlinear Scales}
\shortauthors{Cheng \& Bock}
\graphicspath{{./}{figures/}}

\def\fIHL{$f_{\rm IHL}$}

\begin{document}

\title{Near-infrared Extragalactic Background Light Fluctuations on Nonlinear Scales}

\author[0000-0002-5437-0504]{Yun-Ting Cheng}
\address{California Institute of Technology, 1200 E. California Boulevard, Pasadena, CA 91125, USA}
\email{ycheng3@caltech.edu}

\author{James J. Bock}
\address{California Institute of Technology, 1200 E. California Boulevard, Pasadena, CA 91125, USA}
\address{Jet Propulsion Laboratory, California Institute of Technology, 4800 Oak Grove Drive, Pasadena, CA 91109, USA}

\begin{abstract}
Several fluctuation studies on the near-infrared extragalactic background light (EBL) find an excess power at tens of arcminute scales ($\ell\sim10^3$). Emission from the intra-halo light (IHL) has been proposed as a possible explanation for the excess signal. In this work, we investigate the emission from the integrated galaxy light (IGL) and IHL in the power spectrum of EBL fluctuations using the simulated galaxy catalog MICECAT. We find that at $\ell\sim10^3$, the one-halo clustering from satellite galaxies has comparable power to the two-halo term in the IGL power spectrum. In some previous EBL analyses, the IGL model assumed a small one-halo clustering signal, which may result in overestimating the IHL contribution to the EBL. We also investigate the dependence of the IGL$+$IHL power spectrum on the IHL distribution as a function of redshift and halo mass, and the spatial profile within the halo. Our forecast suggests that the upcoming SPHEREx deep field survey can distinguish different IHL models considered in this work with high significance. Finally, we  quantify the bias in the power spectrum from the correlation of the mask and the signal, which has not been accounted for in previous analyses.
\end{abstract}


\section{Introduction} \label{S:intro}
The extragalactic background light (EBL) arises from the aggregate emission from all sources across cosmic time, and has been measured across the electromagnetic spectrum from gamma rays to the radio. The EBL at different wavelengths originates from different astrophysical sources, and reflects a wide range of physical processes in the universe \citep[see, e.g.,][ for recent reviews]{2016RSOS....350555C,2019ConPh..60...23M}. At near-infrared wavelengths, the integrated galaxy light (IGL) is thought to be primarily produced by redshifted ultraviolet and optical stellar emission from $z\lesssim 2$. However, several EBL absolute photometry measurements \citep{2013PASJ...65..121T,2015ApJ...807...57M,2015ApJ...811...77S,2017ApJ...839....7M,2017NatCo...815003Z,2020ApJ...901..112S,2021ApJ...906...77L,2022AJ....164..170C} have reported excess emission above the IGL \citep{2010ApJ...723...40K,2011MNRAS.410.2556D,2012ApJ...752..113H,2016ApJ...827..108D,2021MNRAS.503.2033K,2021MNRAS.507.5144S}, suggesting potential near-infrared EBL contributions from faint and diffuse populations. EBL fluctuations as a function of wavelength and angular scale probe the EBL sources that trace the underlying matter density field. Near-infrared fluctuation analyses have also reported excess fluctuations over the expected IGL signal at tens of arcminute scales \citep{2005Natur.438...45K,2007ApJ...666..658T,2011ApJ...742..124M,2012ApJ...753...63K,2012Natur.490..514C,2014Sci...346..732Z,2015ApJ...807..140S,2015NatCo...6.7945M,2019PASJ...71...82K,2019PASJ...71...88M}. 

As the scales of excess fluctuations correspond to galaxy nonlinear clustering at low redshifts ($z\lesssim 0.5$), some studies suggest that the excess signal is mostly contributed by stripped stars in dark matter halos, referred to as intra-halo light \citep[IHL; ][]{2012Natur.490..514C,2014Sci...346..732Z,2021ApJ...919...69C}, while others explain it with the emission from first stars and galaxies from the epoch of reionization \citep[EoR; ][]{2002MNRAS.336.1082S,2003MNRAS.339..973S,2005Natur.438...45K,2011ApJ...742..124M,2012ApJ...753...63K,2015NatCo...6.7945M,2021MNRAS.508.1954S}. More exotic candidates responsible for the excess EBL have also been proposed, such as Population III stars and accretion disks around their stellar mass black holes from $z > 7$ \citep{2018ApJS..234...41W}, direct collapse black holes from the dark ages \citep{2013MNRAS.433.1556Y,2014MNRAS.440.1263Y} and the decay of axion-like particles as dark matter candidates \citep[e.g., ][]{2016ApJ...825..104G,2019PhRvD..99b3002K,2021JCAP...05..046C}.

In previous fluctuation analyses \citep[e.g. ][]{2012ApJ...753...63K,2012Natur.490..514C,2014Sci...346..732Z}, the nonlinear clustering from satellite galaxies is based on the analytical halo occupation distribution (HOD) prescription from \citet[][ hereafter H12]{2012ApJ...752..113H}. However, this model adopts the shape of the IGL power spectrum from galaxy number counts, and assumes the galaxy clustering is independent of luminosity, which results in an underestimate of the nonlinear clustering power. \citet{2010ApJ...710.1089F} demonstrate that nonlinear bias from halo emission at the EoR has a significant effect on the EBL power spectrum, which cannot be ignored. Although the galaxy bias is much smaller at lower redshifts than during the EoR, it is still important to investigate the effects of nonlinear clustering from satellite galaxies and the IHL. Therefore, a more realistic modeling at nonlinear clustering scales is essential for interpreting the EBL fluctuation data, particularly for upcoming experiments that will observe EBL fluctuations with high sensitivity and wide spectral coverage such as SPHEREx \citep{2014arXiv1412.4872D, 2018arXiv180505489D}, Euclid \citep{2018LRR....21....2A} and the Roman Space Telescope \citep{2015arXiv150303757S}. For example, SPHEREx deep fields will map the EBL in 200 deg$^2$ across 0.75--5 $\mu$m in 102 spectral channels, promising highly accurate EBL spatial and spectral measurements that will improve constraints on different components in the EBL signal.

In this work, we use the simulated galaxy catalog MICECAT (Sec.~\ref{S:MICECAT}) to model the EBL power spectrum from IGL and IHL at low-redshift ($z \leqslant 1.4$), which captures the bulk of the near-infrared IGL intensity and fluctuations. MICECAT contains information on individual halos and their central and satellite galaxies, quantifying the nonlinear clustering from satellite galaxies, which is ideal for studying the effects on the power spectrum with different IHL models. We investigate the EBL power spectrum with IHL models spanning a wide range of redshift, halo mass, and spatial profile. This allows us to understand the dependence of EBL fluctuations on IHL model parameters and constraints on IHL models with future observations.

This paper is organized as follows. Sec.~\ref{S:Simulations} introduces MICECAT and our application of the catalog to our calculations. Sec.~\ref{S:IGL} presents the MICECAT IGL model and its power spectrum. In Sec.~\ref{S:IHL}, we consider several different IHL models, and analyze their power spectra as a function of the parameters of the model. We summarize our results and discuss contexture outlook and future measurements in Sec.~\ref{S:conclusion}. Throughout this work, we assume a flat $\Lambda$CDM cosmology with $n_s=0.97$, $\sigma_8=0.82$, $\Omega_m=0.26$, $\Omega_b=0.049$, $\Omega_\Lambda=0.69$, and $h=0.68$, consistent with the Planck cosmology \citep{2016A&A...594A..13P}. All fluxes are quoted in the AB magnitude system.

\section{Simulations}\label{S:Simulations}
Our EBL model builds on an $N$-body simulation catalog -- MICECAT (detailed in Sec.~\ref{S:MICECAT}). As it models the three-dimensional distribution of dark matter halos and galaxies, MICECAT contains IGL clustering information on both large (linear) and small (nonlinear) spatial scales. The simulation catalog provides a flexible basis to implement different IGL and IHL models, as well as the effects from observation or data processing such as the instrument point-spread function (PSF) and source masking.

\subsection{MICECAT}\label{S:MICECAT}
We use the MICECAT (v2.0) simulated galaxy catalog \citep{2015MNRAS.448.2987F,2015MNRAS.447.1319F,2015MNRAS.447.1724H}\footnote{For details on MICECAT v2.0, see the documentation: \url{https://www.dropbox.com/s/0ffa8e7463n8h1q/README_MICECAT_v2.0_for_new_CosmoHub.pdf?dl=0}} to model the EBL fluctuations from the IGL and IHL. MICECAT is a product of the $N$-body cosmological simulation Marenostrum Institut de Ci{\'e}ncies de l'Espai Grand Challenge run (MICE-GC), which has 70 billion dark matter particles in a 3072$^3$ Mpc$^3$h$^{-3}$ cubic co-moving box. The dark matter halos are resolved down to $\sim 3 \times 10^{10} M_\odot h^{-1}$.

MICECAT simulates ideal observations of one octant of the sky ($\sim$ 5000 deg$^2$; $0^{\circ}<{\rm R. A.}<90^{\circ}$, $0^{\circ}<{\rm decl.}<90^{\circ}$), and covers the redshift range of $0<z<1.4$ without repeating the simulation box. MICECAT builds on MICE-GC by combining an HOD with subhalo abundance matching to calibrate to observed luminosity functions and clustering \citep{2015MNRAS.447..646C}. MICECAT simulates a mass-limited sample, with deeper coverage in two sub-regions (decl.$>30^{\circ}$; decl.$<30^{\circ}$ and $30^{\circ}<{\rm decl.}<60^{\circ}$), which is complete to $m_i\sim 24$ at all redshifts ($0<z<1.4$). In this work, we only use their simulation in these two deep subregions.

For each galaxy, MICECAT provides its magnitude in several optical and near-infrared bands (SDSS, DES, Euclid, etc), which allows us to model the IGL wavelength dependence. All galaxies in MICECAT have a halo ID and a central/satellite galaxy flag, so we can identify the central galaxy and all satellite galaxies in each halo and use them to model the linear (two-halo) and nonlinear (one-halo) clustering. \citet{2009ApJ...695..900Y} develop an algorithm to group observed galaxies in the Sloan Digital Sky Survey (SDSS) into halos and identify central and satellite galaxies for each halo. As a validation, we compare the conditional stellar mass functions for both central and satellite galaxies in MICECAT and the group catalog from \citet{2009ApJ...695..900Y},  and find that the two catalogs are consistent across halo masses of $10^{12}\lesssim M \lesssim 10^{14}\, M_\odot$.

\subsection{Modeling the EBL with MICECAT}
In this work, we only model the EBL in a single observing band, and leave wavelength dependence studies to future work. We select the Euclid NISP \textit{Y} band from MICECAT, which has an effective wavelength of $\lambda_{\rm eff}=1.07$ $\mu$m. All simulations throughout this work are run on $2\times 2$ deg$^2$ fields selected from the two deep subregions in MICECAT. 
For each field we use $1024\times1024$ pixels, with a $7{''}\times 7{''}$ pixel size, matching CIBER imager data \citep{2013ApJS..207...32B,2014Sci...346..732Z, 2021ApJ...919...69C}. Our choice of field and pixel size corresponds to multipoles $10^2\lesssim\ell\lesssim10^5$, suitable for modeling the nonlinear clustering signal that dominates at tens of arcminute scales ($\ell\sim10^3$). We use 35 such fields (covering 140 deg$^2$ in total) to obtain statistics with a negligible sample variance at the scales of interest. When forecasting sensitivity, we also consider a 200 deg$^2$ survey, the total area of the SPHEREx deep fields near the north and south ecliptic poles.

\begin{figure}[ht!]
\begin{center}
\includegraphics[width=\linewidth]{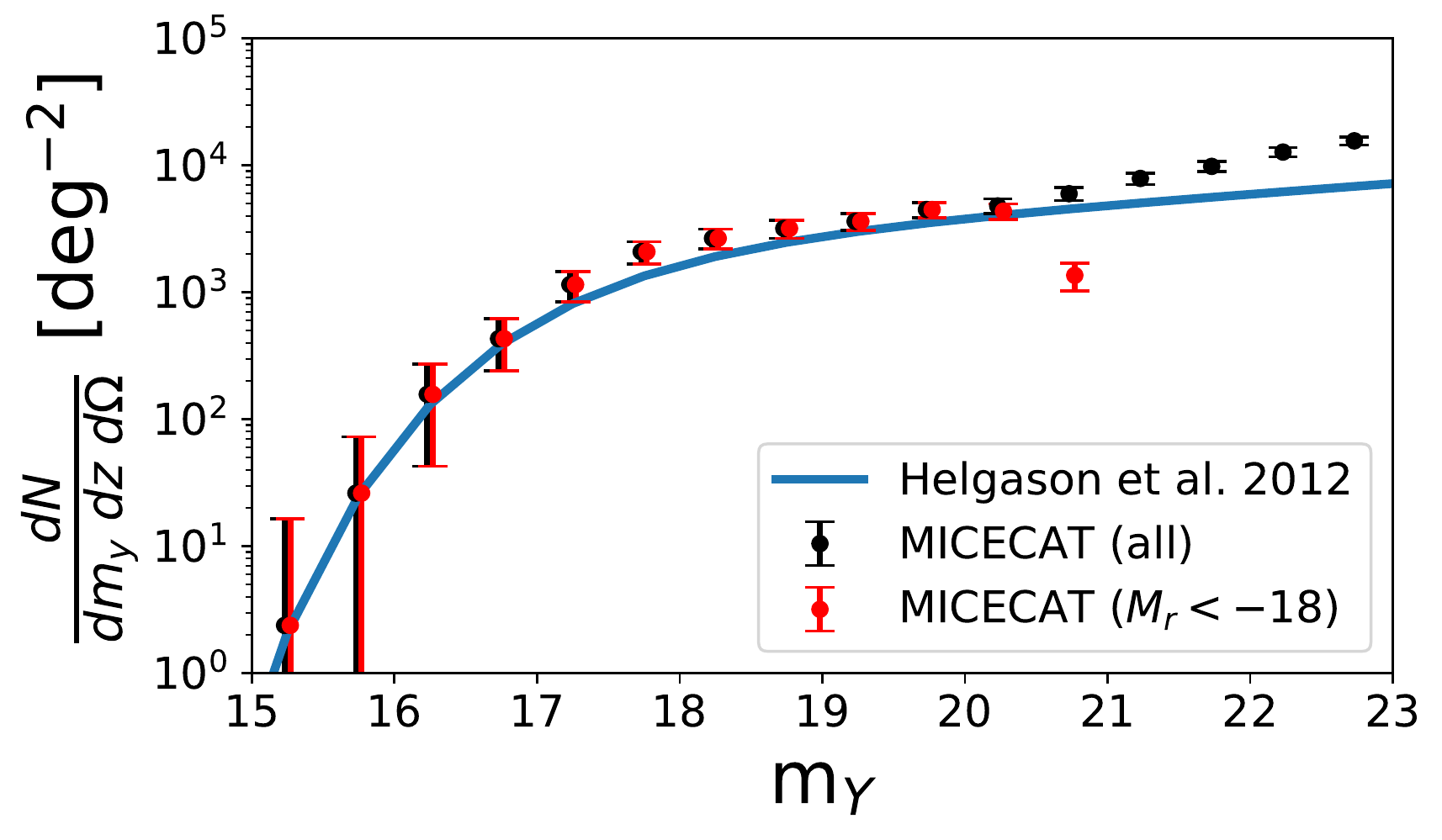}
\caption{\label{F:MICECAT_LF} MICECAT luminosity function at $z=0.2$ in the Euclid NISP \textit{Y} band (black), and the $M_r<-18$ population (red) used in this work. Error bars give the sample variance of the 140 deg$^2$ of MICECAT data we used in this study. The blue line shows a model of luminosity function at the same wavelength and redshift from H12, fitting a Schechter function to observed galaxy counts. The MICECAT luminosity function on the faint end shows an excess above the model, and thus we discard all $M_r>-18$ sources in our analysis.}
\end{center}
\end{figure}

\begin{figure}[ht!]
\begin{center}
\includegraphics[width=\linewidth]{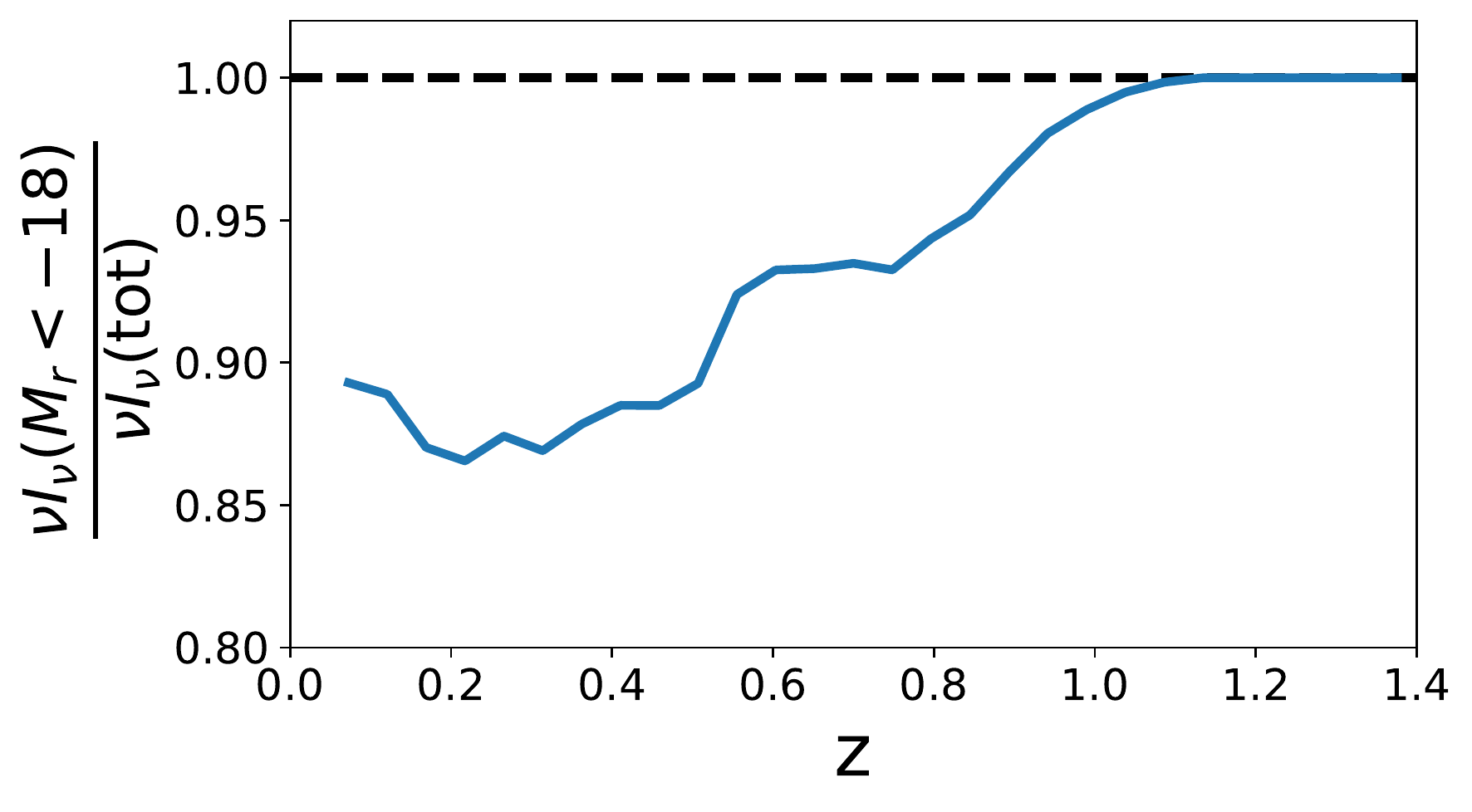}
\caption{\label{F:MICECAT_LF_int} Fraction of the MICECAT IGL from $M_r<-18$ galaxies as a function of redshift. The galaxies fainter than $M_r=-18$ only constitute $\sim 10\%$ of the IGL at $z\lesssim 0.5$, and this fraction decreases to zero at $z\gtrsim 1$.}
\end{center}
\end{figure}

We find an excess number of low-redshift faint galaxies above a Schechter luminosity function \citep{1976ApJ...203..297S} in MICECAT. It is unclear whether these faint sources are consistent with observations or are artifacts from the simulation. H12 shows that galaxy number counts are consistent with the Schechter function fit to the faint end ($m_{\rm AB}\sim 26$ at $\sim1$ $\mu$m), while in \citet{2019ApJ...873...78B} (Fig. 11), the rest-frame $K$ band luminosity function at $z\sim3$ from observation shows a faint-end excess above the Schechter function fit. These sources may be excess and would potentially bias the IGL clustering signal, especially on nonlinear scales as many of them are satellite galaxies in dark matter halos. Furthermore, in reality, faint galaxies are more likely to experience tidal disruptions during the course of their evolution history, which will then be described by the IHL component in our model. Therefore, to make a conservative estimate of the IGL, we discard all $M_r>-18$ MICECAT galaxies, where $M_r$ is the SDSS r-band absolute magnitude given by MICECAT.  Fig.~\ref{F:MICECAT_LF} shows the MICECAT luminosity function at $z=0.2$ in the Euclid NISP \textit{Y} band, and a model of IGL luminosity function from H12, fitted to a Schechter function from observed galaxy counts. We also determined that in the same band, the $M_r>-18$ population only contributes $\sim 10\%$ of the total IGL in MICECAT at $z\lesssim 0.5$; there are almost no $M_r>-18$ sources at $z\gtrsim 1$ (Fig.~\ref{F:MICECAT_LF_int}). In addition, we also checked that discarding $M_r>-18$ sources only reduces $\sim 10\%$ of the IGL power spectrum at $\ell\sim 10^3$. Note that galaxy tidal disruption is a continuous process with no clear boundary between small satellite galaxies and IHL, so we incorporate all $M_r>-18$ galaxies into our model of the IHL component. 

\begin{figure}[ht!]
\begin{center}
\includegraphics[width=\linewidth]{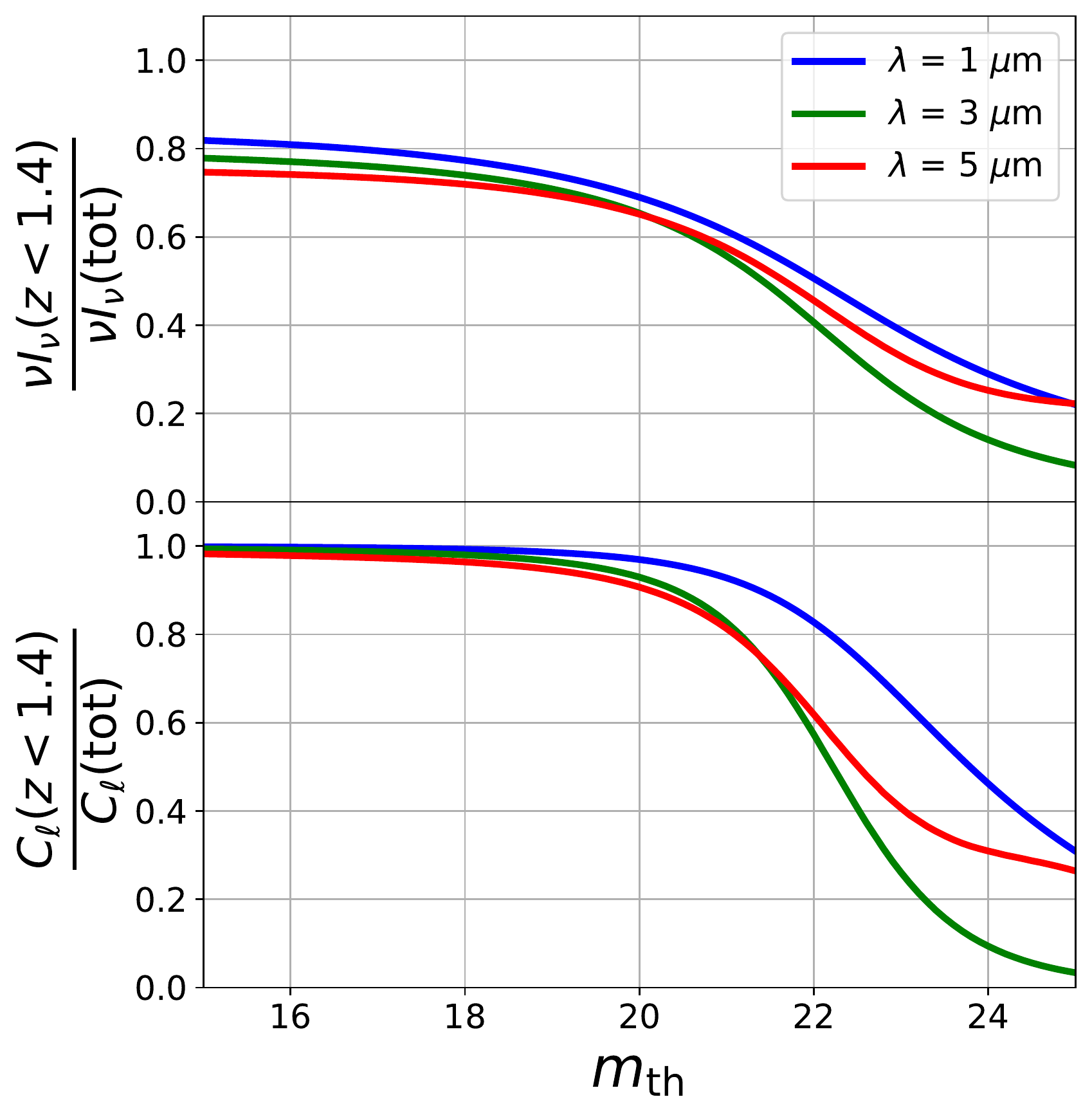}
\caption{\label{F:IClz14_Helgason} Top: fraction of IGL intensity from $z<1.4$ galaxies in H12 model as a function of masking threshold $m_{\rm th}$ at $1$ (blue), $3$ (green), and $5$ (red) $\mu$m. The masking threshold is defined at 1 $\mu$m for all three wavelengths, and the corresponding masking depth at other wavelengths is derived by the abundance matching technique \citep{2022ApJ...925..136C}. Bottom: fraction of IGL power spectrum from linear clustering and Poisson fluctuations at $\ell=10^3$ from $z<1.4$ galaxies. We use the power spectrum model from \citet{2022ApJ...925..136C} built from the H12 IGL model.}
\end{center}
\end{figure}

We use the H12 IGL model to evaluate the EBL signal from higher redshift galaxies that are not included in MICECAT ($z>1.4$). The top panel of Fig.~\ref{F:IClz14_Helgason} shows the fraction of IGL intensity from $z<1.4$ galaxies at three near-infrared wavelengths ($1$, $3$, and $5$ $\mu$m) as a function of flux threshold in AB magnitude $m_{\rm th}$. We consider this fraction with a range of $m_{\rm th}$ at 1 $\mu$m, and infer the masking depth at other wavelengths with the abundance matching recipe from  \citet{2022ApJ...925..136C}. We also compare the IGL two-halo and Poisson power spectrum at $\ell=10^3$ from $z<1.4$ galaxies with the same power spectrum from the total IGL, shown in the bottom panel of\footnote{In Fig.~\ref{F:IClz14_Helgason}, the ratio at $3$ $\mu$m drops faster than the other two wavelengths is because in the H12 model, the $L^*$, characteristic luminosity in the Schechter function, has a maximum at rest-frame $\sim1.5$ $\mu$m, which makes the emission around this wavelength more sensitive to masking. At $z<1.4$, $3$ $\mu$m band emission is dominated by sources around this wavelength, and thus its intensity reduces more by masking than 1 and 5 $\mu$m.} Fig.~\ref{F:IClz14_Helgason}. We use the power spectrum model from \citet{2022ApJ...925..136C}, which is also based on the H12 IGL model. We set $m_{\rm th}=20$ as our fiducial case (see Sec.~\ref{S:masking}), and thus we estimate that at $\sim1$ $\mu$m, the MICECAT catalog captures $\sim 70\%$ and $>95\%$ of the total IGL intensity and power spectrum, respectively. With the cases considered in this work, the higher redshift sources have negligible clustering power at the scales of interest. However, at longer wavelengths or with a deeper masking depth, the IGL will be weighted toward higher redshifts, so MICECAT will not be sufficient to model their signal.

\section{IGL Fluctuations}\label{S:IGL}
We first quantify IGL fluctuations from MICECAT galaxies as point sources, and discuss the effect of PSF broadening in Sec.~\ref{S:mkk}. We decompose the linear and nonlinear clustering signals in MICECAT using a halo model framework \citep[][ Sec.~\ref{S:halo_model}]{2002PhR...372....1C}, and consider masking bright galaxies to increase sensitivity on clustering fluctuations (Sec.~\ref{S:masking}).

\subsection{Halo Model Framework}\label{S:halo_model}
The power spectrum of the MICECAT map consists of clustering and Poisson terms:
\begin{equation}\label{E:Cl}
C_\ell = C_{\ell, {\rm clus}} + C_{\ell, {\rm P}}.
\end{equation}
The clustering power spectrum captures projected large-scale structure fluctuations. Using the Limber approximation,
\begin{equation}\label{E:Cl_clus}
C_{\ell, {\rm clus}} = \int dz \frac{H(z)}{c\,\chi^2(z)} P_{\rm clus}(k, z),
\end{equation}
where $H$, $\chi$, $c$, and $P_{\rm clus}$ are the Hubble parameter, co-moving distance, speed of light, and 3D clustering power spectrum, respectively. The Poisson term arises from the discreteness of galaxies:
\begin{equation}\label{E:Cl_shot}
C_{\ell, {\rm P}} = \frac{1}{\Omega_{\rm sur}}\sum_i \left( \nu F_{\nu,i}\right)^2,
\end{equation}
where $\Omega_{\rm sur}$ is the survey angular size, $F_{\nu,i}$ is the specific flux of the $i$th galaxy, and we sum over all galaxies in the map.

In the halo model framework, $P_{\rm clus}$ can be decomposed into two-halo (linear) and  one-halo (nonlinear) terms, $P_{\rm 2h}$ and $P_{\rm 1h}$, which describe the correlations between different halos and within the same halo, respectively:
\begin{equation}\label{E:P_clus}
P_{\rm clus}(k,z) =P_{\rm 1h}(k,z) + P_{\rm 2h}(k,z),
\end{equation}
and we can also express the angular power spectrum in terms of these two components:
\begin{equation}
C_{\ell, {\rm clus}} =C_{\ell, {\rm 1h}} + C_{\ell, {\rm 2h}},
\end{equation}
where
\begin{equation}\label{E:Cl_1h}
C_{\ell, {\rm 1h}} = \int dz \frac{H(z)}{c\,\chi^2(z)} P_{\rm 1h}(k, z),
\end{equation}
and 
\begin{equation}\label{E:Cl_2h}
C_{\ell, {\rm 2h}} = \int dz \frac{H(z)}{c\,\chi^2(z)} P_{\rm 2h}(k, z).
\end{equation}
The one- and two-halo 3D power spectrum can be expressed as\footnote{We drop the redshift symbol $z$ hereafter for clarity.}
\begin{equation}\label{E:P_1h}
P_{\rm 1h}(k) =\frac{1}{\bar{n}_g^2}\int dM \frac{dn}{dM} I^2_h(M) u^2(k|M),
\end{equation}
and
\begin{equation}\label{E:P_2h}
P_{\rm 2h}(k) =P_{\rm lin}(k)\left [\frac{1}{\bar{n}_g}\int dM \frac{dn}{dM} I_h(M)b_h(M) u(k|M)\right ]^2,
\end{equation}
where $\bar{n}_g$ is the mean number density of galaxies, $\frac{dn}{dM}$ is the halo mass function, $I_h(M)$ and $b_h(M)$ are the mean intensity and bias from all sources within a mass $M$ halo, respectively, $P_{\rm lin}$ is the linear matter power spectrum, and $u(k|M)$ is the Fourier transform of the normalized intensity radial profile of mass $M$, which approaches $1$ on large scales (small $k$), and rolls off at the scale of the halo. We directly evaluate $C_{\ell, {\rm 1h}}$ and $C_{\ell, {\rm 2h}}$ from the MICECAT catalog instead of performing the integration in Eq.~\ref{E:P_1h} and \ref{E:P_2h}, described below.

As each galaxy in MICECAT has a halo ID, we can calculate the one- and two-halo correlation functions from all galaxy pairs within the same halo and across different halos, respectively. The power spectra are the Fourier transform of the two correlation functions. Calculating the correlation function is computationally expensive, as it requires $O(N^2)$ operations for $N$ galaxies. Therefore, we use an approximate estimator to derive one- and two-halo power spectra directly from 2D maps as follows. First, we make an intensity map of all the central and satellite galaxies in MICECAT (hereafter, referred to as the \textit{full map}). The power spectrum of the full map gives the total clustering signal $C_{\ell, {\rm clus}}$ that contains both one- and two-halo terms. Next, we make another map by moving all satellite galaxies to the same location as their central galaxies (hereafter, referred to as the \textit{central map}). We define the power spectrum of the central map as our two-halo power spectrum estimator, $\hat{C}_{\ell, {\rm 2h}}$. The one-halo power spectrum estimator is defined as $\hat{C}_{\ell, {\rm 1h}} = C_{\ell, {\rm clus}} - \hat{C}_{\ell, {\rm 2h}}$.

Our central map does not contain one-halo correlations, and its halo profile is a Dirac delta function as all fluxes are from the halo center, and thus $u(k|M)=1$ for all $k$. Therefore, $\hat{C}_{\ell, {\rm 1h}}$ and $\hat{C}_{\ell, {\rm 2h}}$ is given by the integration of $\hat{P}_{\rm 1h}$ and $\hat{P}_{\rm 2h}$ with Eq.~\ref{E:Cl_1h} and \ref{E:Cl_2h}, respectively, where
\begin{equation}\label{E:P_2h_hat}
\hat{P}_{\rm 2h}(k) =P_{\rm lin}(k)\left [\frac{1}{\bar{n}_g}\int dM \frac{dn}{dM} I_h(M)b_h(M) \right ]^2,
\end{equation}
and
\begin{equation}\label{E:P_1h_hat}
\hat{P}_{\rm 1h}(k) = P_{\rm clus}(k) - \hat{P}_{\rm 2h}(k).
\end{equation}

We calculate $C_{\ell, {\rm clus}}$ and $\hat{C}_{\ell, {\rm 2h}}$ from the power spectra of the full and central-only map, respectively. From Eq.~\ref{E:P_2h} and \ref{E:P_2h_hat}, we note that on large scales where $u(k|M)\sim 1$, $\hat{P}_{\rm 2h}(k) \approx P_{\rm 2h}(k)$, our estimator converges to the exact expression. On small scales, $\hat{P}_{\rm 2h}(k) > P_{\rm 2h}(k)$, and thus $\hat{P}_{\rm 1h}(k) < P_{\rm 1h}(k)$, so our one/two-halo power spectrum estimator is a lower/upper bound of the true power spectrum, respectively. 

The scales at which $\hat{C}_{\ell,{\rm 2h}}$ deviates from $C_{\ell,{\rm 2h}}$ are the angular size of halos. We estimate this scale by stacking halos in MICECAT. With our fiducial masking threshold $m_{\rm th}=20$ (Sec.~\ref{S:masking}), the stacked halo profile from satellite galaxies has an FWHM $\sim 15{''}$ ($\ell\sim4\times10^4$, black dashed line in Fig.~\ref{F:IGL_Cl}). Therefore, at the scales of interest (tens of arcminute scales), the error from our one- and two-halo power spectrum approximation assuming $u(k|M)=1$ is negligible. As an additional check, we randomly select $10\%$ of the galaxies in a $2\times2$ deg$^2$ MICECAT field, and compute the two-point correlation function of the two-halo term in the full map and the central map. The results of the two maps are consistent at $\gtrsim 100$ arcsec ($\ell\lesssim 6500$), and thus the error from our approximated two-halo term is negligible on the scales of interest. We estimate the Poisson term from the high-$\ell$ power spectrum, where the clustering signal is negligible, and we subtract Poisson fluctuations from the total power spectrum to get the clustering terms.

\subsection{Masking}\label{S:masking}
We consider masking bright galaxies before measuring the power spectrum. With simulations, we can perform a ``perfect masking'' by removing bright source fluxes from the IGL map. In reality, masking is usually implemented by removing pixels near bright sources. Pixel masking leaves residual flux outside the masking region that depends on the PSF, and will also cause mode coupling in the power spectrum. We set the fiducial masking scheme to perfect masking with an AB magnitude threshold $m_{\rm th}=20$ in the Euclid NISP \textit{Y} band. All the results shown in this paper are using this fiducial masking scheme if not otherwise specified. The effect of the PSF and pixel masking will be discussed in Sec.~\ref{S:mkk}.

\subsection{IGL Power Spectrum}\label{S:IGL_PS}
\begin{figure}[ht!]
\begin{center}
\includegraphics[width=\linewidth]{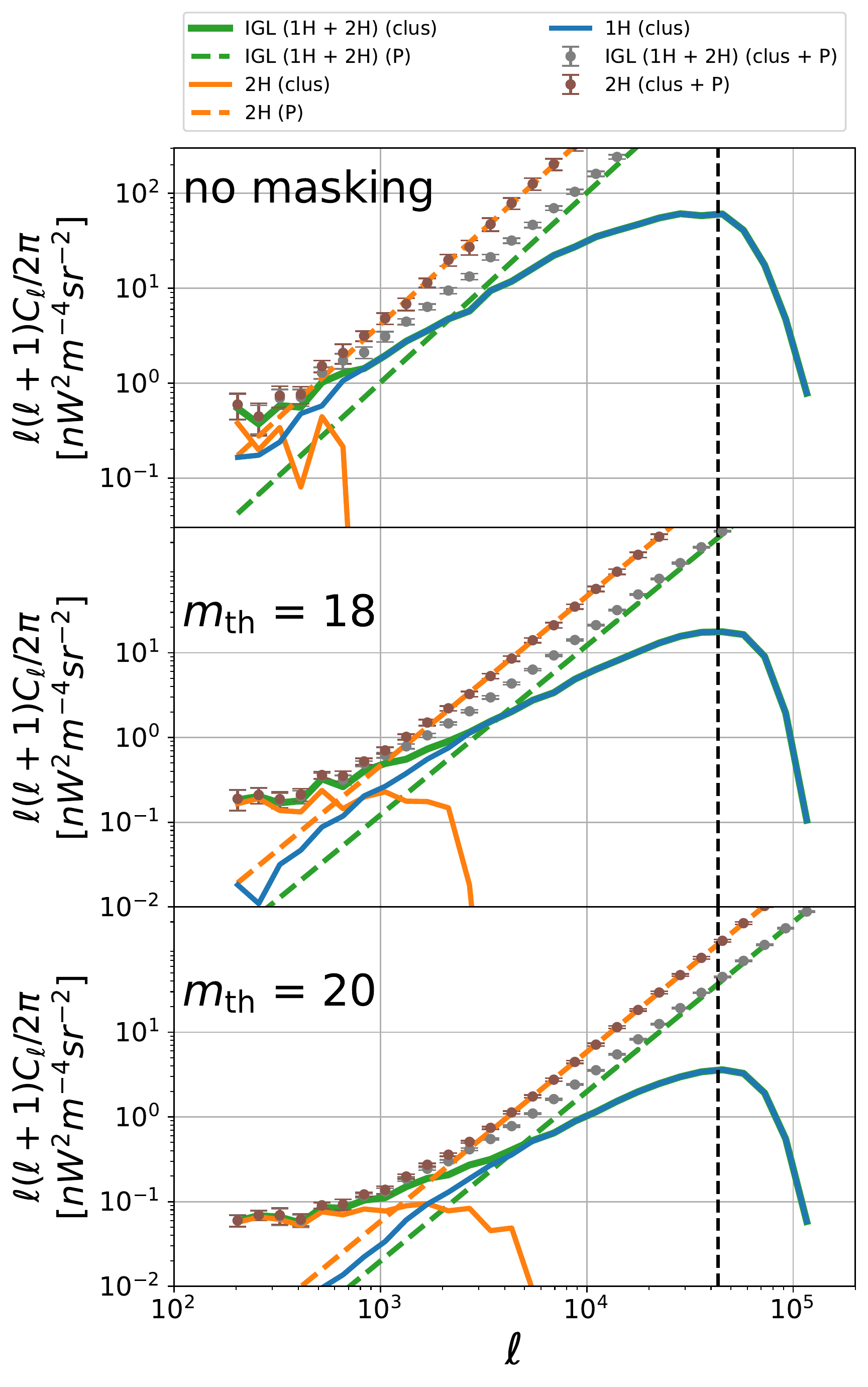}
\caption{\label{F:IGL_Cl}The IGL power spectrum from MICECAT with no masking (top) and with a ``perfect'' masking threshold $m_{\rm th}=18$ (middle) and $20$ (bottom). Gray data points give the power spectrum of the total IGL. Brown data points show the two-halo power spectrum estimator from the central map. Error bars in all cases give the projected uncertainty in a 200 deg$^2$ observation. The green and orange dashed lines give Poisson fluctuations in the full and central maps, respectively, which are fitted at high $\ell$. Green and orange solid lines give the clustering power spectrum of the full and central maps, respectively, defined by the difference between the total power spectrum and the Poisson term. The blue line gives the one-halo clustering estimator, which is the difference between the total clustering signal (green line) and the two-halo clustering estimator (orange line). The black dashed line is the angular scale of the average halo profile from stacking, which also corresponds to the turnover $\ell$ mode for the one-halo power spectrum. Our approximation of one- and two-halo power spectra ($\hat{C}_{\ell, {\rm 1h}}$ and $\hat{C}_{\ell, {\rm 2h}}$) is valid for scales larger than this dashed line.}
\end{center}
\end{figure}

Fig.~\ref{F:IGL_Cl} presents the components of the IGL power spectrum from MICECAT with no masking, $m_{\rm th}=20$ masking (fiducial case), and $m_{\rm th}=18$ masking (similar to the masking depth adopted in \citet{2014Sci...346..732Z}). The one- and two-halo components are derived from the full and central maps (see Sec.~\ref{S:halo_model}).

\begin{figure}[ht!]
\begin{center}
\includegraphics[width=\linewidth]{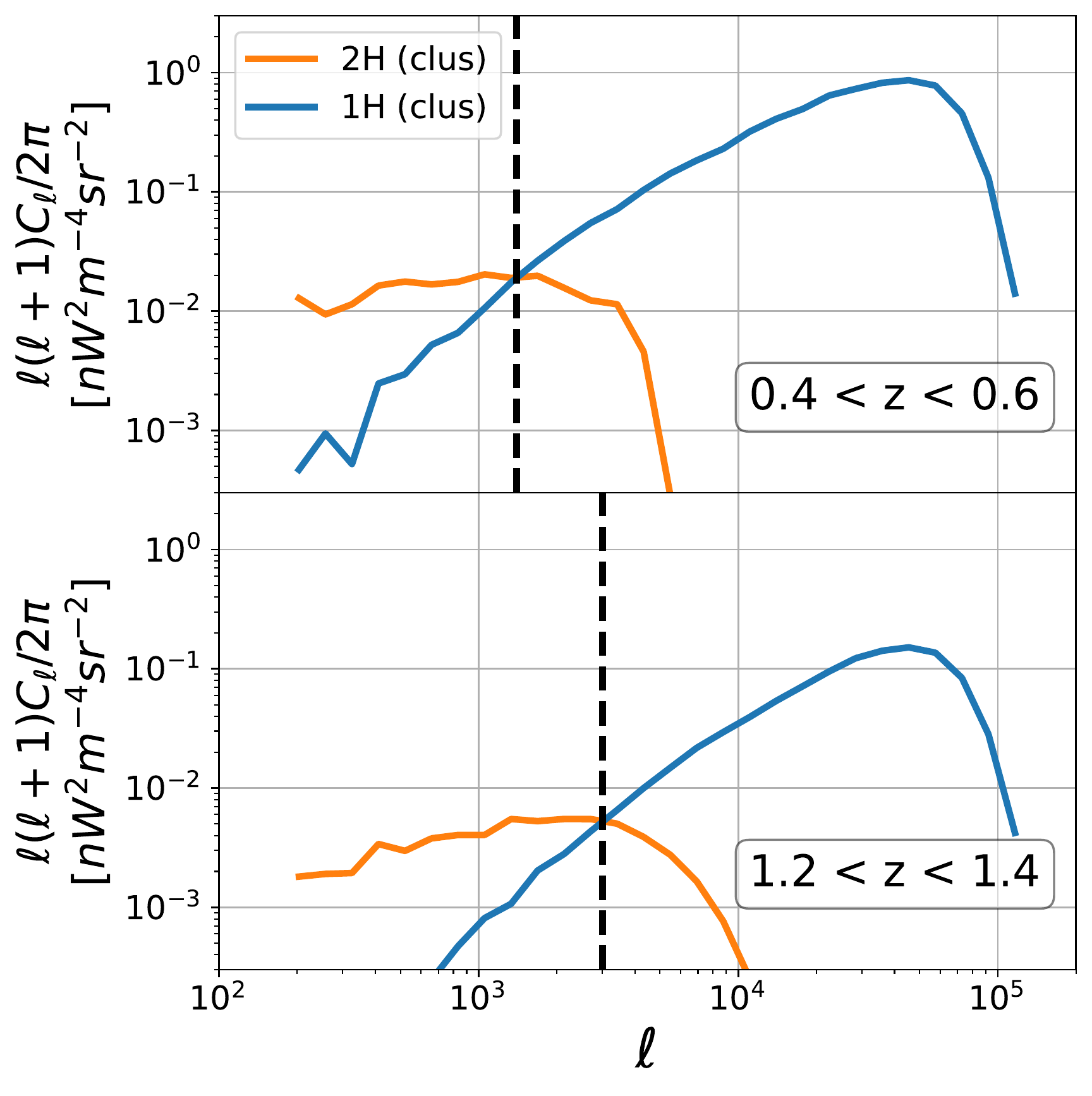}
\caption{\label{F:IGL_Cl_z} The one-(blue) and two-halo (orange) clustering power spectrum from MICECAT with sources at $0.4<z<0.6$ (top) and $1.2<z<1.4$ (bottom) with masking threshold $m_{\rm th}=20$. The crossover angular scale of the two terms (black dashed lines) is smaller at higher redshifts.}
\end{center}
\end{figure}

With deeper masking, the clustering signal is dominated by smaller halos, and thus the crossover of one- and two-halo power spectrum occurs at larger $\ell$ modes. We note that the crossover scale also depends on the redshift and halo mass distribution of $f_{\rm IHL}$. If there is more IHL emission from more massive or lower redshift halos, the average angular size of IHL will be larger and thus the crossover will be at smaller $\ell$ modes. Fig.~\ref{F:IGL_Cl_z} compares the MICECAT power spectrum in redshift slices $0.4<z<0.6$ and $1.2<z<1.4$ at $m_{\rm th}=20$, where we observe the crossover scale of one- and two-halo terms is larger at lower redshift.

In both cases with masking in Fig.~\ref{F:IGL_Cl}, we see that one- and two-halo terms have comparable power at $\sim 10{'}$ scales ($\ell\sim 10^3$). This indicates that when modeling EBL fluctuations at this scale, the non-linear clustering from IGL needs to be properly handled. In some previous fluctuation analyses \citep[e.g., ][]{2012ApJ...753...63K,2012Natur.490..514C,2014Sci...346..732Z}, the IGL model is adopted from H12, which uses the HOD model fitted to the SDSS galaxy number counts. This model assumes galaxy clustering is independent of galaxy luminosity, which results in a fixed shape in the IGL (one- and two-halo) power spectrum, where different wavelengths and masking limits only affect the overall amplitude. Furthermore, for the power spectrum of the IGL intensity field studied here, the contribution from each galaxy is weighted by its flux, whereas in the case of SDSS galaxy clustering, all galaxies have equal weights in the power spectrum. These simplifications in H12 lead to a small one-halo term that equals the two-halo power at $\sim4'$ ($\ell\sim 2500$). For example, as shown in Fig.~\ref{F:IGL_Cl}, at the masking threshold used in the analysis of \citet[][$m_{\rm th}=18$]{2014Sci...346..732Z}, our model predicts a larger contribution from the one-halo term (the one- and two-halo power spectrum crossover at $\ell\sim800$), implying that the nonlinear clustering from satellite galaxies in this previous analysis is underestimated, and thus their excess power has to be explained by other sources such as the IHL.

\section{Intra-Halo Light}\label{S:IHL}
Next, we add IHL to our model. We first define the IHL fraction, $f_{\rm IHL}$, as the fractional added luminosity from IHL to the halo. In the near infrared, we consider only IGL and IHL contributions to the total halo emission, so 
\begin{equation}
f_{\rm IHL} = \frac{F_{\nu,{\rm IHL}}}{F_{\nu,{\rm IGL}}+F_{\nu,{\rm IHL}}}.
\end{equation}
In reality, $f_{\rm IHL}$ will depend on redshift, halo mass, and the time evolution of the halo. Different models give different IHL properties in each halo, resulting in a different scale-dependent contribution to the one- and two-halo IGL+IHL power spectra. For simplicity, we set our fiducial IHL model to the case where all halos have the same $f_{\rm IHL}$ regardless of their redshift and mass, and assume the IHL spatial distribution follows a Navarro–Frenk–White (NFW) profile \citep{1996ApJ...462..563N}. In the following subsections, we apply different redshift, mass, and spatial profile dependence to the fiducial model, and discuss their impact on the power spectrum.

\begin{figure}[ht!]
\begin{center}
\includegraphics[width=\linewidth]{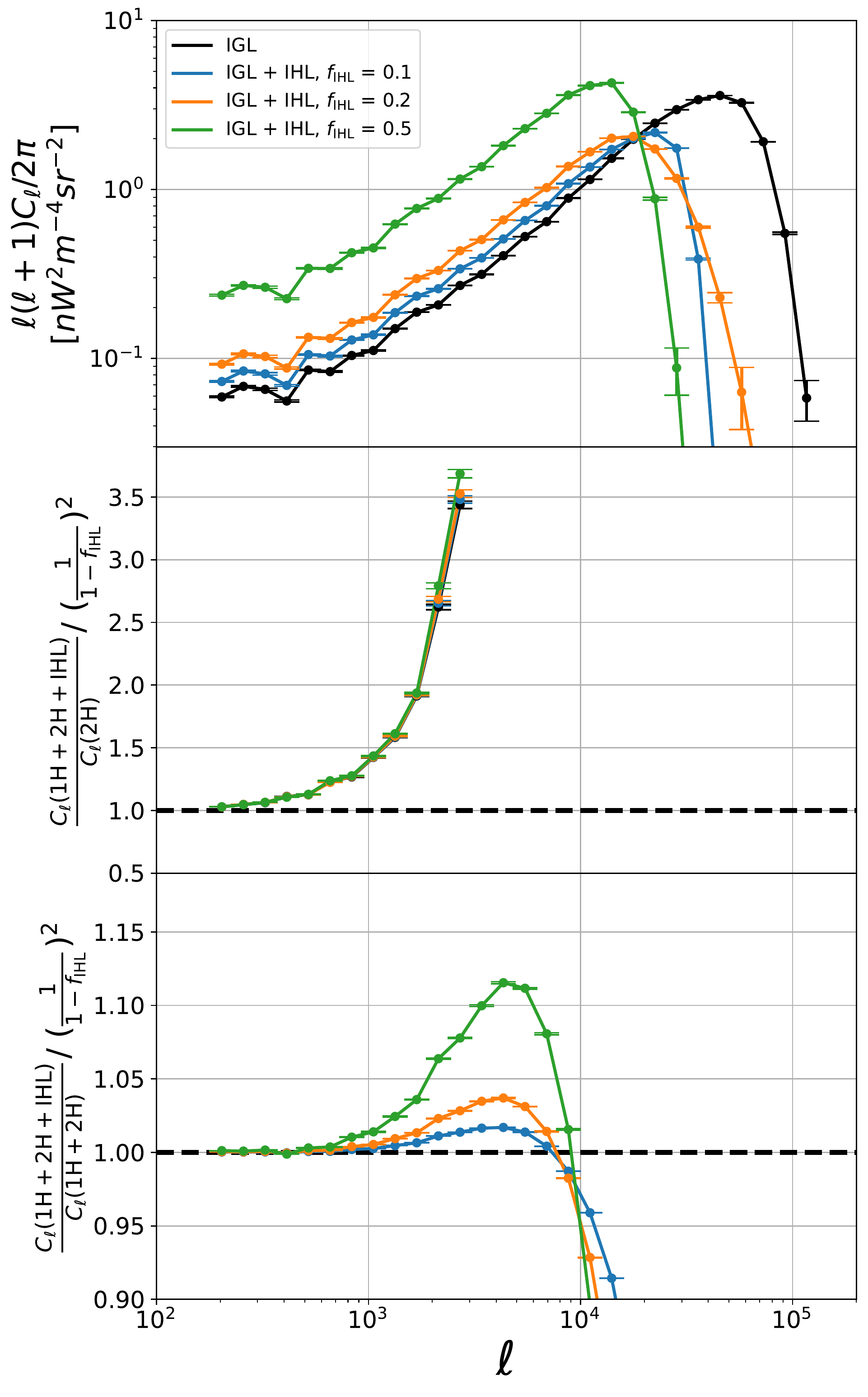}
\caption{\label{F:IHL_Cl} Top: clustering power spectrum of IGL$+$IHL (one-halo $+$ two-halo) with IHL fractions $f_{\rm IHL}=0$ (black), $0.1$ (blue), $0.2$ (orange), and $0.5$ (green). Here, we show the fiducial case where all halos have the same $f_{\rm IHL}$, and the IHL follows an NFW profile. Error bars denote sample variance for a 200 deg$^2$ survey. Middle: ratio of the IGL$+$IHL power spectrum to the two-halo IGL power spectrum, normalized by $\left(\frac{1}{1-f_{\rm IHL}}\right)^2$. We only show the low-$\ell$ modes where there is sufficient sensitivity on the two-halo power spectrum. The one-halo power in the IHL model closely follows the one-halo power in the MICECAT simulation with small departures at $\ell\sim 5000$ indicated below. Bottom: ratio of the IGL$+$IHL and IGL (one-halo $+$ two-halo) power spectra of the same three $f_{\rm IHL}$ cases. The normalization factor $(\frac{1}{1-f_{\rm IHL}})^2$ accounts for the change in the two-halo amplitude with $f_{\rm IHL}$.}
\end{center}
\end{figure}

We first investigate the dependence of the clustering power spectrum on $f_{\rm IHL}$ values in the fiducial case (Fig.~\ref{F:IHL_Cl}). On scales larger than the typical halo size ($\ell\sim5\times10^3$), both IGL (one-halo $+$ two-halo) and IHL trace the matter density field, so the IGL$+$IHL power spectra converge to the pure two-halo IGL power spectrum (as the one-halo term is suppressed on large scales) scaled by the square of the total intensity $\frac{F_{\nu,{\rm IGL}}+F_{\nu,{\rm IHL}}}{F_{\nu,{\rm IGL}}}=(\frac{1}{1-f_{\rm IHL}})^2$. Both satellite galaxies (one-halo term) and the IHL produce excess power at the halo scale. With our fiducial IHL model, the IHL gives a slightly higher excess signal compared to the satellite galaxies (see bottom panel of Fig.~\ref{F:IHL_Cl}), where their difference in the power spectrum is $\sim 10\%$ if $f_{\rm IHL}=0.5$, and less than $5\%$ with two other lower $f_{\rm IHL}$ cases ($f_{\rm IHL}=0.1, 0.2$) at the halo scale. Our model indicates that in a 200 deg$^2$ survey limited by sample variance, this difference in the power spectrum can be detected with high significance in all three $f_{\rm IHL}$ values considered here.

\citet{2014Sci...346..732Z} measured fluctuations with the CIBER instrument, reporting excess fluctuation power over the IGL on angular scales $500<\ell<2000$. However, as seen in the middle panel of Fig.~\ref{F:IHL_Cl}, both MICECAT and the IHL model predict a significant one-halo contribution to the total power spectrum, increasing the total power by factors between 1.2 and 2.2 on these scales. Accounting for the one-halo contribution thus reduces the inferred IHL luminosity. 

We tried fitting the CIBER 1.1 $\mu$m power spectrum measurements to our fiducial IGL$+$IHL model, but our constant $f_{\rm IHL}$ model does not accurately fit the shape of the measured power spectrum. The overall amplitude is set by points at $\ell > 10^4$, which have high statistical weight, and the fixed shape then underpredicts the measured power at $\ell < 3000$. This suggests that our fiducial model assuming a constant $f_{\rm IHL}$ for all halos is not a good description to the data, and thus a more sophisticated model that includes redshift and halo mass dependence or other free parameters is essential for building a realistic IHL model for fluctuation measurements, which we leave to future work.
% \yt{\citet{2014Sci...346..732Z} measured fluctuations with the CIBER instrument, reporting excess fluctuation power over the IGL on angular scales $500<\ell<2000$. However, as seen in the middle panel of Fig.~\ref{F:IHL_Cl}, both MICECAT and the IHL model predict a significant one-halo contribution to the total power spectrum, increasing the total power by 1.2 to 2.2 on these scales. Accounting for the one-halo contribution thus reduces the inferred IHL luminosity. If we assume a band centered at $\ell=1000$, when the one-halo contribution increases the total power by 1.4, the inferred IHL luminosity in \citet{2014Sci...346..732Z} drops from 7.0 nW m$^{-2}$ sr$^{-1}$ to 4.4 nW m$^{-2}$ sr$^{-1}$ at 1.1 $\mu$m, and from 11.4 nW m$^{-2}$ sr$^{-1}$ to 8.2 nW m$^{-2}$ sr$^{-1}$ at 1.6 $\mu$m.}

\subsection{Redshift Dependent \fIHL}\label{S:IHL_z}
\begin{figure}[ht!]
\begin{center}
\includegraphics[width=\linewidth]{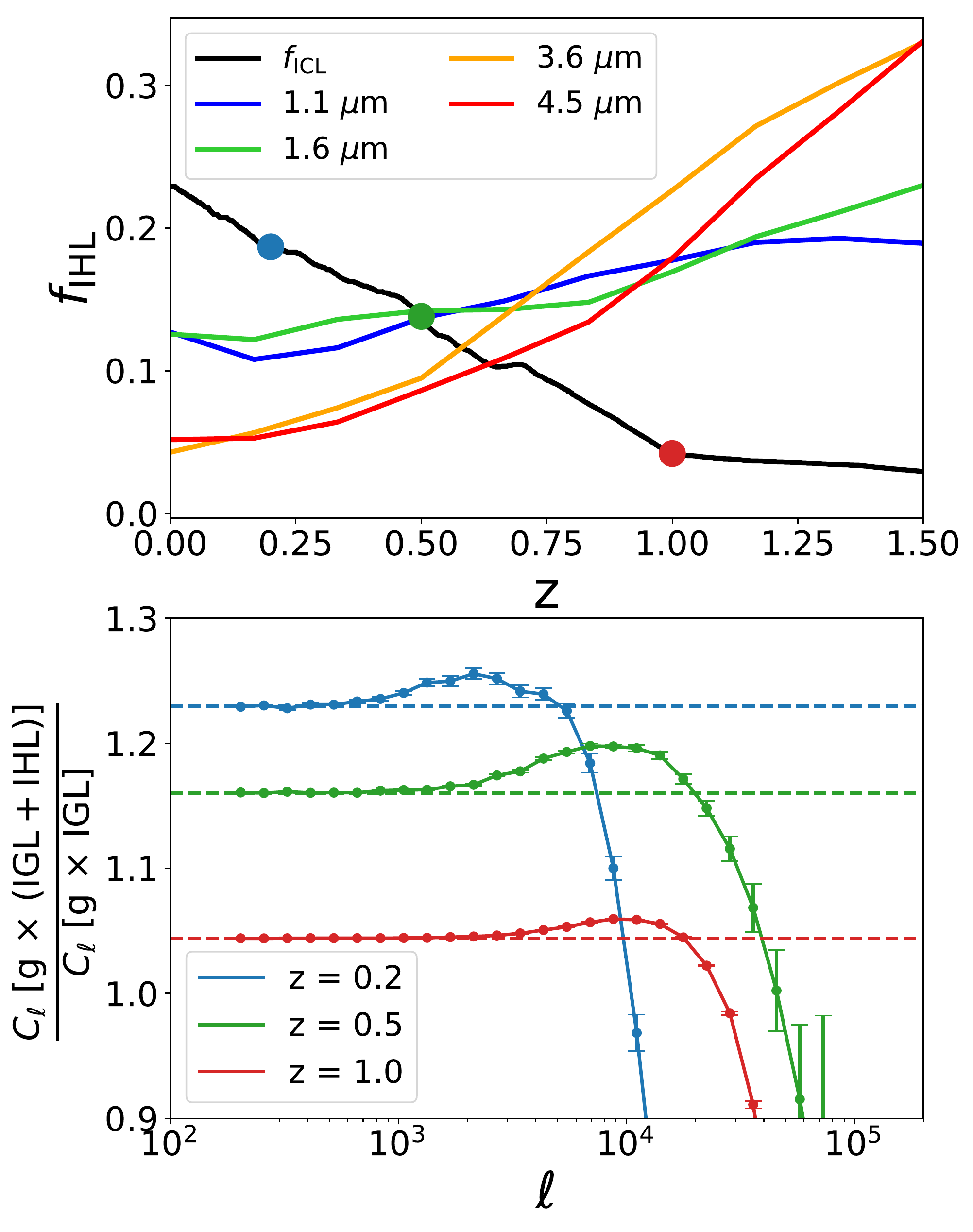}
\caption{\label{F:IHL_xcorr} Top: our model of the $f_{\rm IHL}$--$z$ relation, adopting the empirical ICL fraction ($f_{\rm ICL}$) from \citet{2018MNRAS.479..932C} (black). The IHL model from \citet{2012Natur.490..514C} in four near-infrared bands is also shown for comparison (colored lines). The three colored dots correspond to the redshift of the power spectra shown in the bottom panel. Bottom: ratio of galaxy-IGL$+$IHL and galaxy-IGL cross power spectra at $z=0.2$ (blue), $0.5$ (green), and $1$ (red). Error bars denote sample variance in a 200 deg$^2$ survey. 
The dashed lines are  the ratio of IGL$+$IHL and IGL flux at each redshift, the expected ratio $\frac{1}{1-f_{\rm IHL}}$ on large scales.}
\end{center}
\end{figure}

W apply a redshift dependence to our $f_{\rm IHL}$ model while using the same fixed mass dependence and the NFW profile as the fiducial model. Therefore, in our model, the majority of the IHL emission is associated with the characteristic sizes of halos ($\sim 10^{12}\, M_\odot$ at $z=0$) that dominate the IGL emission. Although it is difficult to observe IHL in typical-sized halos individually due to their low surface brightness, some model-dependent constraints have been placed from EBL fluctuation measurements \citep{2012Natur.490..514C,2014Sci...346..732Z}. Fig.~\ref{F:IHL_xcorr} shows the $f_{\rm IHL}$--$z$ relation from \citet{2012Natur.490..514C}, where their model is fitted to 3.6 and 4.5 $\mu$m fluctuation maps from Spitzer. In addition, the same model in two near-infrared bands in CIBER (1.1 and 1.6 $\mu$m) is also shown in Fig.~\ref{F:IHL_xcorr}. \citet{2014Sci...346..732Z} study the near-infrared fluctuations with CIBER, and constrain the IHL using a model similar that of \citet{2012Natur.490..514C}. For comparison, Fig.~\ref{F:IHL_xcorr} also shows the stellar mass fraction of the intra-cluster light (ICL) from \citet{2018MNRAS.479..932C} (their ``STANDARD'' model). We see that in \citet{2012Natur.490..514C} model, the $f_{\rm IHL}$ increases with redshift for $z<1.5$, and the redshift dependence is stronger for longer wavelengths; whereas \citet{2018MNRAS.479..932C} suggest that for massive galaxy clusters, $f_{\rm ICL}$ decreases with redshift. While there are very limited observational constraints on the IHL in a typical-sized halo, and the dependence of $f_{\rm IHL}$ on redshift and halo mass is highly degenerate in EBL fluctuation measurements, more sensitive observations on individual halos or numerical simulations are required to explain the $f_{\rm IHL}$ color dependence and the opposite redshift trend in the IHL and the IGL. As a demonstration in this work, we adopt the $f_{\rm ICL}$ from \citet{2018MNRAS.479..932C} for all halos at a given redshift as our model for the $f_{\rm IHL}$--$z$ relation.

Cross correlating the 2D EBL images with galaxies at known redshift \citep{2022ApJ...925..136C} enables redshift tomography, since EBL emission only correlates with external tracers from the same redshift. In Fig.~\ref{F:IHL_xcorr} we also show a forecast of cross correlation with the SPHEREx galaxy catalog, with a  projected number density given by their public products\footnote{\url{https://github.com/SPHEREx/Public-products/blob/master/galaxy_density_v28_base_cbe.txt}}, using photometric redshift errors of $0.01<\sigma_z/(1+z)<0.03$. To implement cross correlation with MICECAT, we consider redshift slices with width $\Delta z=0.03\times(1+z)$, and generate a tracer galaxy catalog by choosing the same number of brightest galaxies as the SPHEREx number density in the Euclid $Y$ band. The cross power spectrum between the IGL$+$IHL intensity map and the tracer catalog at $z=0.2$, $0.5$,  and $1$ shown  in the bottom panel of Fig.~\ref{F:IHL_xcorr} readily detects the redshift dependence of $f_{\rm IHL}$ as well as the subtle effect of the NFW profile in the IHL model compared with one-halo clustering.

\subsection{Mass Dependent \fIHL}\label{S:IHL_M}
\begin{figure}[ht!]
\begin{center}
\includegraphics[width=\linewidth]{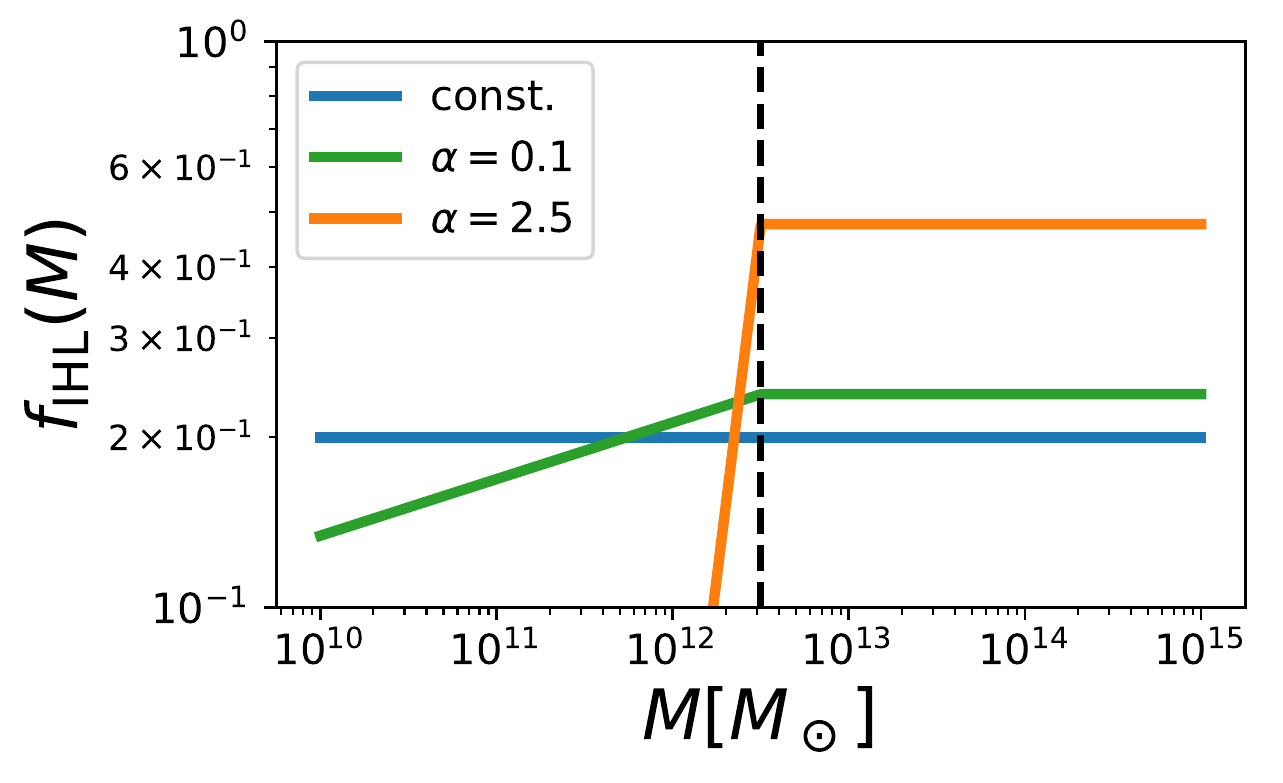}
\caption{\label{F:f_IHL_M} Three models of the $f_{\rm IHL}$--$M$ relations considered in this work. The constant model (blue) is our fiducial case where all the halos have the same IHL fraction. The other two models are a power law with slope $\alpha=2.5$ (orange) and $0.1$ (green), respectively, for $M<10^{12.5}\,M_\odot$, and extending the value at $10^{12.5}\,M_\odot$ (black dashed line) to all $M>10^{12.5}\,M_\odot$ halos. The three curves are normalized such that the mean IHL fraction $\bar{f}_{\rm IHL}=0.2$, where $\bar{f}_{\rm IHL}$ is defined as the fraction of total EBL flux from the IHL.}
\end{center}
\end{figure}

Next, we investigate the effect of a halo mass dependence on $f_{\rm IHL}$. Fig.~\ref{F:f_IHL_M} shows the three $f_{\rm IHL}$--$M$ relations considered in this work. The first model is the fiducial case where $f_{\rm IHL}$ is independent of halo mass. In the second and the third model, we use a power law with slope $\alpha=2.5$ and $0.1$ (i.e., $f_{\rm IHL}(M)\propto M^{2.5}$ and $f_{\rm IHL}(M)\propto M^{0.1}$), respectively, for $M<10^{12.5}\,M_\odot$, and extending the value at $10^{12.5}\,M_\odot$ to all $M>10^{12.5}\,M_\odot$ halos. The $\alpha=2.5$ model approximates the $f_{\rm IHL}(M)$ function from an analytical model at $z=0$ \citep{2007ApJ...666...20P, 2008MNRAS.391..550P}, whereas $\alpha=0.1$ is the best-fit power-law slope found in \citet{2014Sci...346..732Z}, who fit the $f_{\rm IHL}$--$M$ relation from the near-infrared fluctuations in CIBER images. To compare the three cases, we normalize the $f_{\rm IHL}(M)$ function to the same mean IHL fraction, $\bar{f}_{\rm IHL}=0.2$, where $\bar{f}_{\rm IHL}$ is defined as the fraction of total EBL flux from the IHL, and we only consider IGL and IHL as the source of total EBL emission.

\begin{figure*}[ht!]
\begin{center}
\includegraphics[width=\linewidth]{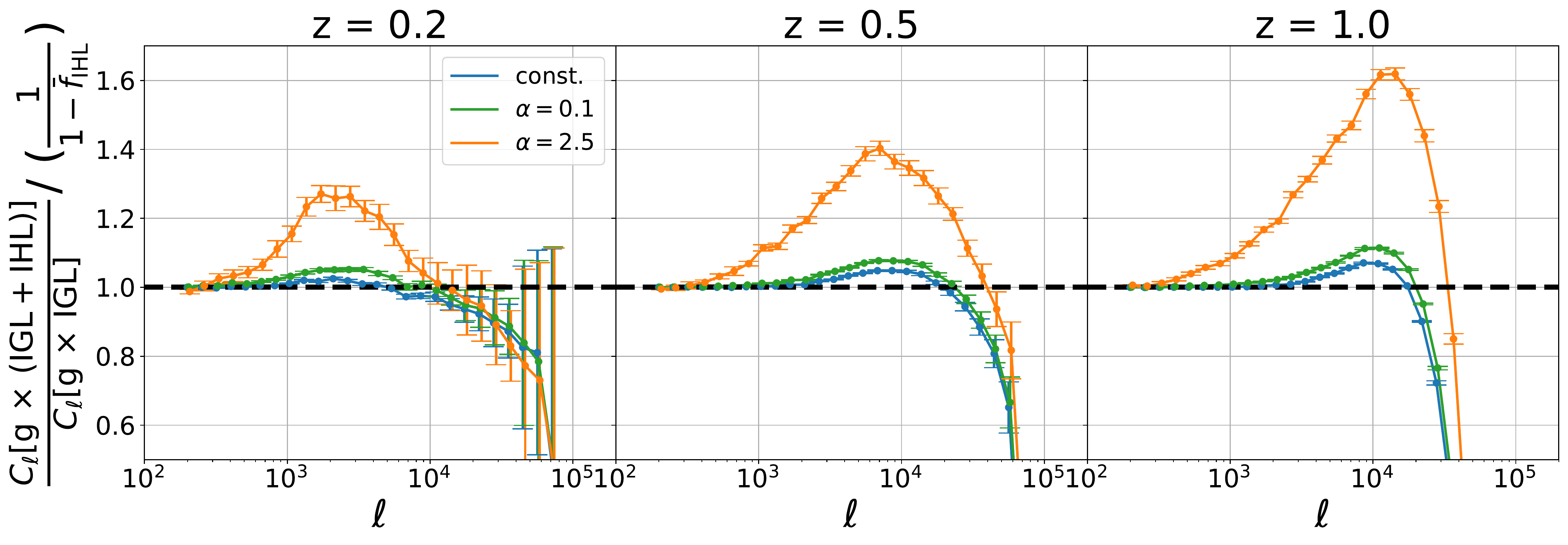}
\caption{\label{F:Cl_f_IHL_M} Ratio of the galaxy-IGL$+$IHL and the galaxy-IGL cross power spectrum of the three $f_{\rm IHL}(M)$ models shown in Fig.~\ref{F:f_IHL_M} at $z=0.2$ (left),  $0.5$ (middle), and $1$ (right). Error bars are the sample variance in a 200 deg$^2$ survey. The values are scaled by the mean ratio of the IGL$+$IHL and the IGL flux ($\frac{1}{1-\bar{f}_{\rm IHL}}$), which is the expected ratio on large scales. Here, we use $\bar{f}_{\rm IHL}=0.2$.}
\end{center}
\end{figure*}

We calculate the ratio of the galaxy-IGL$+$IHL and the galaxy-IGL cross power spectra for the three $f_{\rm IHL}(M)$ models at $z=0.2$,  $0.5$, and $1$, as shown in Fig.~\ref{F:Cl_f_IHL_M}. We consider the same redshift slices and the same galaxy catalog as Sec.~\ref{S:IHL_z}. The ratio is scaled by $\frac{1}{1-\bar{f}_{\rm IHL}}$, which is the expected scaling of two-halo clustering on large scales. Our model suggests that with the same $\bar{f}_{\rm IHL}$ value, the mass dependence of $f_{\rm IHL}$ has a significant impact on fluctuations at cluster angular scales. In the $\alpha=2.5$ case, the majority of the IHL emission is from massive halos, which results in large fluctuation power on one-halo scales compared to the other two cases.

\subsection{Power Spectrum with Different IHL Profiles}\label{S:IHL_profile}

\begin{figure*}[ht!]
\begin{center}
\includegraphics[width=\linewidth]{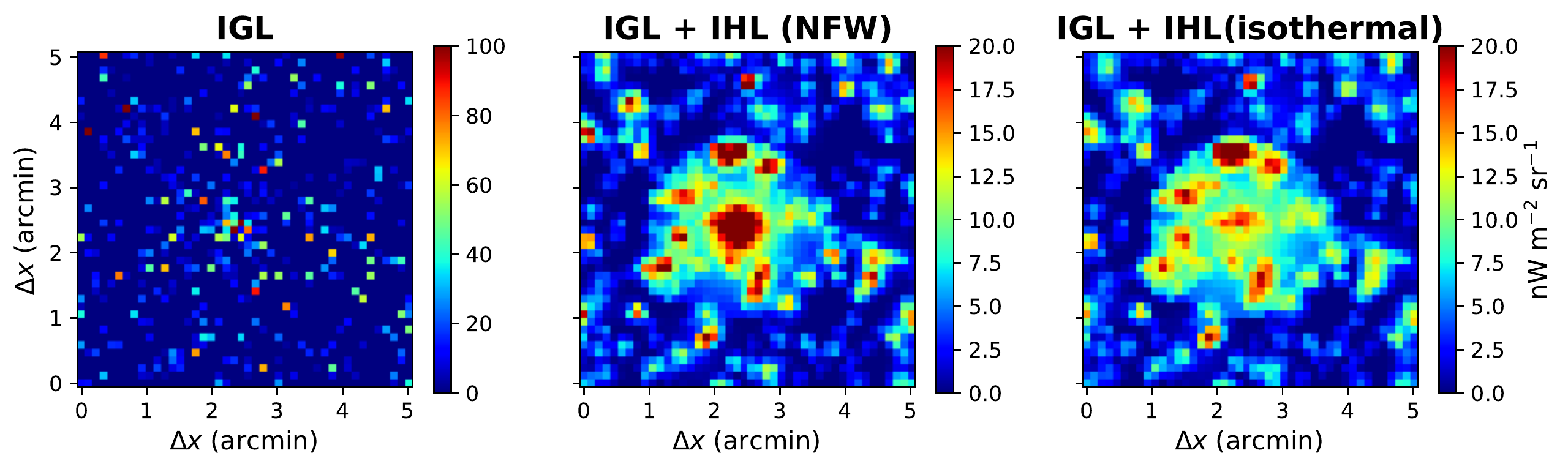}
\caption{\label{F:IHL_prof_img} An example of a $5'\times5'$ image gridded in our $7{''}\times7{''}$-sized pixel. We choose a region centered at a galaxy cluster in MICECAT. Left: IGL intensity field.
Middle: IHL$+$IGL field with fiducial IHL model that follows an NFW profile. Left: IHL$+$IGL field with fiducial IHL model but using isothermal halo profile. The isothermal model has a more extended halo profile than the NFW model.}
\end{center}
\end{figure*}

\begin{figure}[ht!]
\begin{center}
\includegraphics[width=\linewidth]{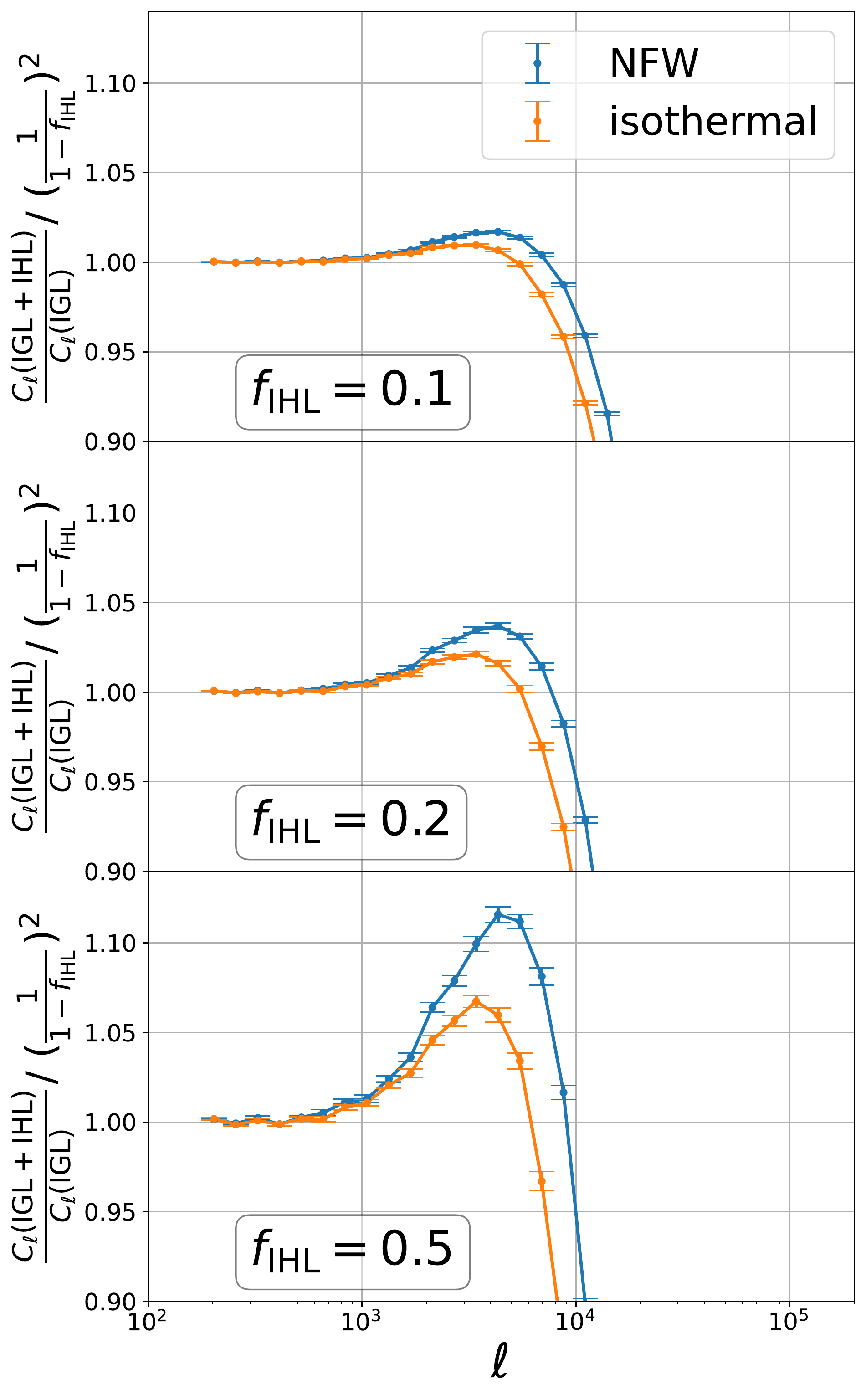}
\caption{\label{F:IHL_prof_Cl} Ratio of IGL$+$IHL and IHL power spectra using NFW (blue) and isothermal (orange) IHL profiles with $f_{\rm IHL}=0.1$ (top), $0.2$ (middle), and $0.5$ (bottom). We assume $f_{\rm IHL}$ is independent of redshift and mass as in our fiducial model. Error bars are the sample variance in a 200 deg$^2$ survey. We normalize the ratio by the factor $(\frac{1}{1-f_{\rm IHL}})^2$ (the ratio of the square of total intensity), which is the expected value on large scales.}
\end{center}
\end{figure}

We compare the power spectrum from different IHL profiles while using constant $f_{\rm IHL}$--$z$ and $f_{\rm IHL}$--$M$ relations in our fiducial model. We consider an isothermal halo profile, which has a more extended light distribution compared to the fiducial NFW profile shown in Fig.~\ref{F:IHL_prof_img}. In Fig.~\ref{F:IHL_prof_Cl} we show the ratio of the IGL$+$IHL and the IGL auto power spectra for the two models. In a 200 deg$^2$ survey, our forecast shows that the two IHL profile models can be distinguished with high significance. In reality, IHL is formed from tidal disruption interactions, and thus the average distribution given in an intensity mapping observation could be more complicated than the two simple models considered here. While Fig.~\ref{F:IHL_prof_Cl} is based on auto-spectra, measurements based on galaxy redshift cross spectra will also be useful in studying IHL profiles.

\subsection{Effects of the PSF and Masking}\label{S:mkk}
In practice, masking to remove pixels contaminated by sources is always imperfect due to the broadening effect of the PSF. The PSF and pixel masks result in two systematic effects: (1) residual fluxes outside the mask (hereafter, referred to as \textit{PSF halo}) introduce extra intensity and correlations at the scale of mask size, and (2) pixel masks not only make the observed image highly irregular but are also correlated with unmasked sources \citep[see, e.g.,][]{2021MNRAS.503.5310R}. Hereafter, this effect is referred to as \textit{mask-signal correlation}. Both factors introduce mode coupling to the power spectrum. Therefore, the power spectrum estimation will be biased if these effects are not properly taken into account in the mode-coupling correction procedure.

\begin{figure*}[ht!]
\begin{center}
\includegraphics[width=\linewidth]{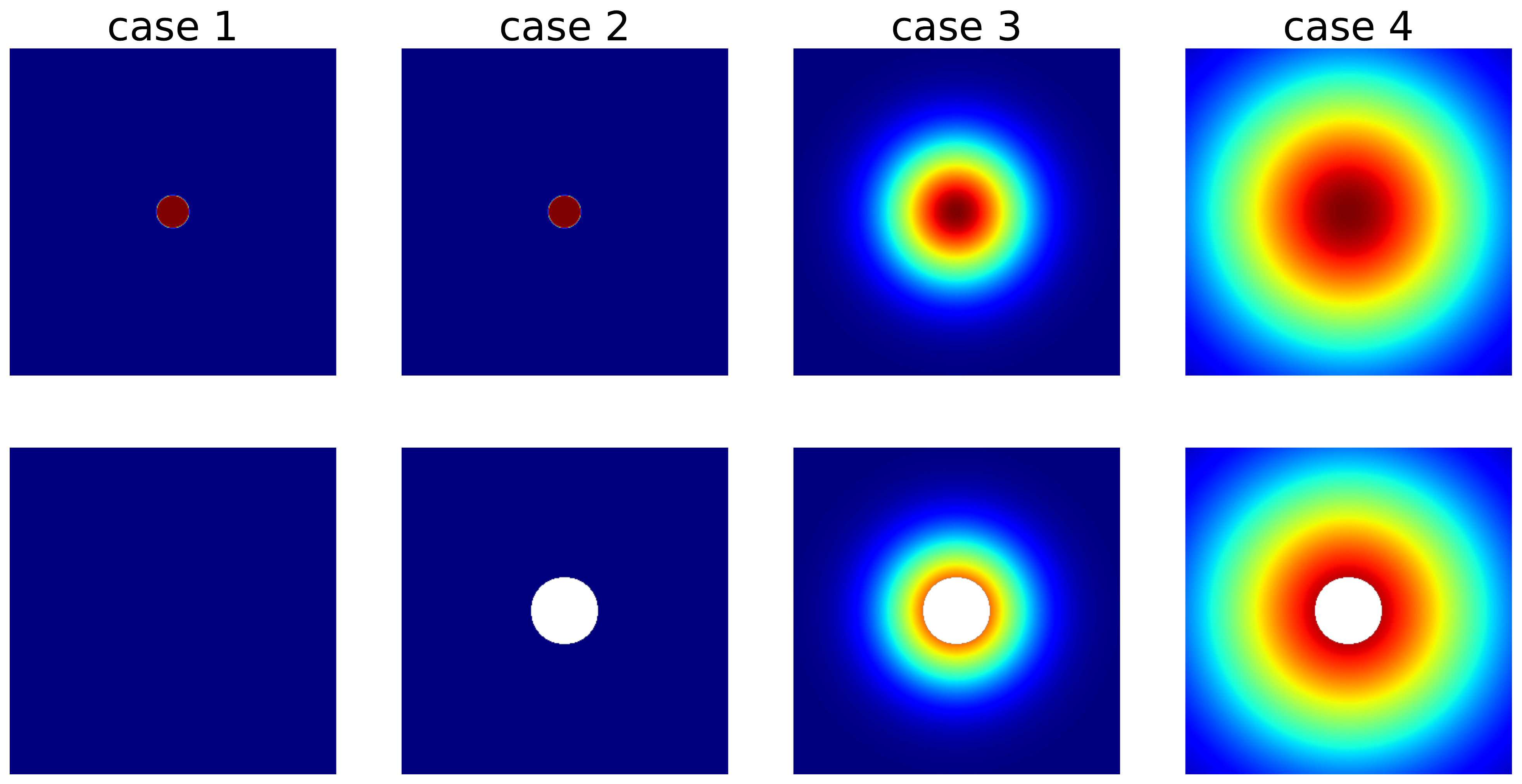}
\caption{\label{F:masking_cases}Illustration of the four masking cases in Sec.~\ref{S:mkk}. The top and bottom rows show a source on an image before and after masking, respectively. In cases 1 and 2, there is no PSF so the source is a point source on the image; in cases 3 and 4, we consider a smaller and a larger Gaussian PSF. Case 1 implements perfect masking that completely removes the flux from the source, whereas for the other three cases, we use pixel masking and the white circle denotes the masking region. Cases with a PSF leave low-level residual halos that extend beyond the mask.}
\end{center}
\end{figure*}

To estimate these systematic effects, we incorporate a Gaussian PSF into the MICECAT simulated images and then mask pixels. We implement the mode-coupling correction algorithm from \citet{2014Sci...346..732Z}, where the mode-coupling matrix is derived by simulating the response from pure-tone Gaussian random fields. However, in these simulations, the mask and the signal are not correlated. Furthermore, residual flux from imperfect masking (PSF halos) can introduce a bias. We investigate the resulting bias with these simulations. We compare the power spectrum of four different cases:
\begin{enumerate}
    \item Fiducial case (case 1): no PSF ($\sigma_{\rm PSF}=0$); perfect masking (i.e., the masking is performed by removing bright sources). In this case, there is no systematic error from either the mask-signal correlation or PSF halos.
    \item Pixel masking-only case (case 2): no PSF ($\sigma_{\rm PSF}=0$); pixel masking. In this case, there is systematic effect from the mask-signal correlation but no PSF halos.
     \item Pixel masking and $\sigma_{\rm PSF}=7{''}$ case (case 3): Gaussian PSF with $\sigma_{\rm PSF}=7{''}$ (size of one pixel); pixel masking. In this case, there are systematic errors from both mask-signal correlation and PSF halos.
     \item Pixel masking and $\sigma_{\rm PSF}=70{''}$ case (case 4): same as case 3 but with $\sigma_{\rm PSF}=70{''}$, which exaggerates both effects.
\end{enumerate}
The four cases are illustrated in Fig.~\ref{F:masking_cases}. In case 2--4, we use the same mask and the same mode-coupling matrix. The masks are circles centered at sources with a magnitude-dependent radius, and we mask all galaxies brighter than 18 AB magnitude in the Euclid $Y$ band with the radius--magnitude relation from \citet{2014Sci...346..732Z}.

\begin{figure*}[ht!]
\begin{center}
\includegraphics[width=\linewidth]{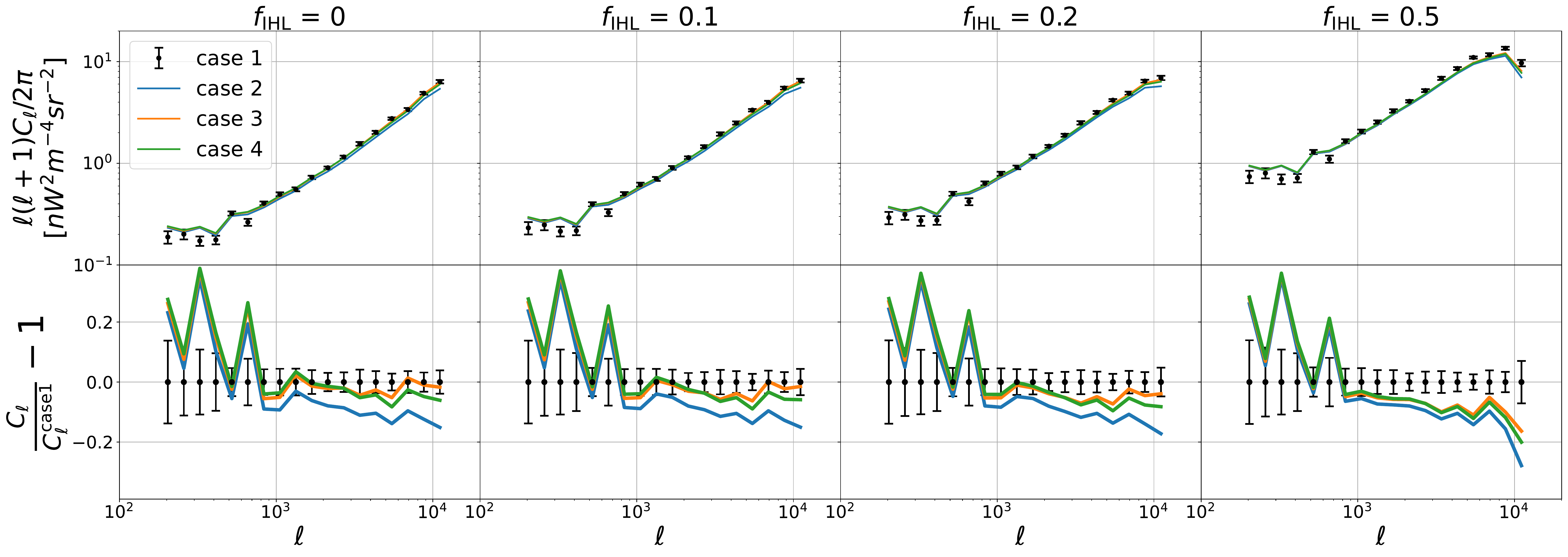}
\caption{\label{F:mkk_Cl} Top: power spectra of the four cases described in Sec.~\ref{S:mkk} with four $f_{\rm IHL}$ values. Black: fiducial case (case 1); blue: pixel masking-only case (case 2); orange: pixel masking and $\sigma_{\rm PSF}=7{''}$ case (case 3); green: pixel masking and $\sigma_{\rm PSF}=70{''}$ case (case 4). Error bars are the sample variance in a 200 deg$^2$ survey. Bottom: fractional error with respect to the fiducial case.}
\end{center}
\end{figure*}

Fig.~\ref{F:mkk_Cl} compares the power spectrum of the four cases described above with different $f_{\rm IHL}$ values. The power spectrum of the pixel masking-only case (case 2) is systematically lower than the fiducial case (case 1) below one-halo scales ($\ell\sim 10^3$) due to the unaccounted mask-signal correlation. The two cases with a non-negligible PSF (cases 3 and 4) have higher power than the pixel masking-only case at all scales due to the PSF halos. On large scales ($\ell\lesssim 10^3$), the power spectrum is proportional to the square of total intensity, and therefore the $\sigma_{\rm PSF}=70{''}$ case has the highest power because it has more residual flux outside the mask. On small scales, the power spectrum of the $\sigma_{\rm PSF}=70{''}$ case is lower than the $\sigma_{\rm PSF}=7{''}$ case due to the uncorrected PSF roll-off. For all four example scenarios considered here, our simulation suggests that the mask-signal correlation and PSF halos only cause a $\sim10\%$ bias on the power spectrum. However, we note that the bias from these systematic effects will in the general case depend on the PSF, masking radius, masking depth, mode-coupling correction algorithm, and the underlying clustering of the signal.

\section{Conclusion}\label{S:conclusion}
In this work, we modeled the near-infrared EBL from the IGL and IHL using the numerical simulation MICECAT, focusing on fluctuation signals at non-linear clustering scales. The MICECAT galaxy catalog provides the galaxy--halo relation for quantifying IGL non-linear clustering from satellite galaxies. We also modeled the signal from the IHL, and investigated the power spectrum dependence on different IHL models. Finally, we considered the effects of the PSF and pixel masking, and showed the systematic bias on the power spectrum if these effects are not properly taken into account. In summary:
\begin{enumerate}
    \item  The one-halo clustering from satellite galaxies and the two-halo clustering have comparable power at tens of arcminute scales ($\ell\sim10^3$). Previous studies have found excess fluctuations at these scales in near- and mid-infrared, and explain the excess signal with emission from the EoR \citep{2012ApJ...753...63K} or the low-redshift IHL \citep{2012Natur.490..514C,2014Sci...346..732Z}. However, we find their IGL models underpredict the nonlinear clustering from satellite galaxies, which can account for some of the excess power.
    \item We investigated the IHL power spectrum with different redshift, halo mass, and IHL profile dependence. Our forecast shows that with a 200 deg$^2$ survey, such as the SPHEREx deep field, the models considered in this work can be distinguished with high significance, assuming that sample variance is the dominant source of uncertainty.
    \item We simulated the systematic bias on the power spectrum from the mask-signal correlation and residual bright source fluxes in the presence of PSF and pixel masking. As a demonstration, we apply the same masking and mode-coupling correction procedure as \citet{2014Sci...346..732Z} to the MICECAT simulated map. Our results show that the systematic bias from these effects is modest in our example case. In practice, this bias depends on the specifics of the mask, PSF,  and the mode-coupling correction algorithm, as well as the underlying clustering of the signal.
 \end{enumerate}

While we only explored some simple IHL models in this work, our results show that upcoming data from SPHEREx promises to probe detailed properties of the IHL from EBL fluctuation measurements. Here, we detail a few future directions toward more sophisticated signal models and analysis tools to further improve IGL and IHL constraints.

First, it is crucial to develop a detailed model of the IGL and IHL that describes their dependence on redshift, halo mass, and spatial profile. In this work, we explore this dependence separately by only varying one factor in each test case. In reality, the IHL properties as a function of redshift, halo mass, and spatial profile are likely to be a more complicated relation rather than a separable function. In addition, since the IHL arises from stripped stars in halos, its properties and evolution history is tightly related to the IGL. Therefore, it is essential to model the IGL and IHL in a coherent manner. Hydrodynamic simulations that simultaneously capture large-scale cosmological clustering and the small-scale physics of the IHL and satellite galaxies will provide useful models for future EBL measurements.

Furthermore, our current model based on MICECAT is limited to $z<1.4$, which is not sufficient to model the signal at longer wavelengths ($\sim3-5$ $\mu$m) where $z\sim2$ galaxies dominate the IGL and IHL emission.

Modeling the wavelength dependence of the IGL and IHL and other EBL components is also an essential building block for analyzing data from future experiments. For example, SPHEREx will carry out a low-resolution spectroscopy of the EBL from $0.75-5$ $\mu$m. The large number of available band auto and cross correlations provide a new means to study both the EBL redshift history and source spectra, so it is essential to have a detailed model of the EBL component wavelength and redshift dependence.

In addition, to prepare for upcoming EBL observations,  it is also important to explore optimal statistics to extract the desired information from the data. For example, in Sec.~\ref{S:IHL_z}, we discuss that cross correlation will be a powerful tool to study the redshift dependence of the EBL. Besides redshift tomography, we can also cross-correlate EBL maps with tracers selected by a certain brightness or galaxy type to infer the IGL and IHL associated with those tracers.

\begin{acknowledgments}

We are grateful to Mike Zemcov for providing analysis on refitting our model to CIBER data. We thank Asantha Cooray, Tzu-Ching Chang, and the CIBER and SPHEREx collaborations for helpful discussions. We thank Mike Zemcov, Shuji Matsuura, and Richard Feder for valuable comments that improve the manuscript. This work was supported by NASA grants NNX16AJ69G and 80NSSC20K0595. This work has made use of CosmoHub. CosmoHub has been developed by the Port d'Informaci{\'o} Cient{\'i}fica (PIC), maintained through a collaboration of the Institut de F{\'i}sica d'Altes Energies (IFAE) and the Centro de Investigaciones Energ{\'e}ticas, Medioambientales y Tecnol{\'o}gicas (CIEMAT) and the Institute of Space Sciences (CSIC \& IEEC), and was partially funded by the "Plan Estatal de Investigaci{\'o}n Científica y Técnica y de Innovación" program of the Spanish government.

\end{acknowledgments}

\vspace{5mm}
\software{
astropy \citep{2013A&A...558A..33A,2018AJ....156..123A}
}

% \appendix

\bibliography{IHL}{}
\bibliographystyle{aasjournal}

\end{document}